# Testing the link between terrestrial climate change and Galactic spiral arm transit

Spiral arms and terrestrial climate

Andrew C. Overholt[1], Adrian L. Melott[1], and Martin Pohl[2]

ABSTRACT

We re-examine past suggestions of a close link between terrestrial climate change and the Sun's transit of spiral arms in its path through the Milky Way galaxy. These links produced concrete fits, deriving the unknown spiral pattern speed from terrestrial climate correlations. We test these fits against new data on spiral structure based on CO data that does not make simplifying assumptions about symmetry and circular rotation. If we compare the times of these transits to changes in the climate of Earth, not only do the claimed correlations disappear, but also we find that they cannot be resurrected for any reasonable pattern speed.

[1] Department of Physics and Astronomy, University of Kansas, 1251 Wescoe Dr. # 1082, Lawrence, KS 66045, USA; andrewo@ku.edu, melott@ku.edu

[2] Department of Physics and Astronomy, Iowa State University, Ames, IA 50011-3160, USA;  mkp@iastate.edu

1. INTRODUCTION

In recent years there have been suggestions of a strong correlation between spiral-arm passages of the Sun in its orbit around the Galaxy and changes in the terrestrial climate. This connection has been based on a statistical association of spiral-arm passages with the timing of ice ages as well as the abundance of $^{18}O$ in fossils, which is linked to the ocean water isotopic ratio. The $^{18}O$ in the ocean is enriched when $^{16}O$ water preferentially evaporates and is locked up in ice. The presumed mechanism for the climate change is thought to be an increase in cosmic rays that may affect cloud formation.

As the solar system passes through the spiral arms of our galaxy, it should be exposed to an increased rate of cosmic rays from more frequent regional supernovae and stronger cosmic-ray confinement on account of a higher amplitude of magnetic turbulence. These protons interact in our atmosphere creating many secondary particles that have been suggested in past works to play a role in climate variations. One model suggests that these secondary particles may be responsible for an increased number of low clouds being formed, thus blocking sunlight and cooling the climate (Svensmark 2006). Previous work on this subject has suggested that these spiral-arm crossings and the resultant increase in cosmic rays would therefore decrease the overall temperature of the planet (Shaviv 2003, Shaviv & Veizer 2003). This suggested mechanism is controversial, both on grounds of the weak response of cloud-condensation nuclei to changes in the cosmic-ray flux (e.g. Pierce & Adams 2009), and on geological grounds (Rahmstorf et. al. 2004), but the controversy continues: see Erlykin et al. (2009) and Svensmark et al. (2009). We do not address this mechanism here, nor conflicting geological arguments, but rather investigate whether the correlation between spiral-arm passage and climate holds up under new information on the structure of the galaxy.

Since the primary publications on the spiral-arm passage and climate correlation, newer models of the galactic structure, including the positions of the spiral arms, have been created that place our transit of these spiral arms at a different time. Models of the galactic structure have been formulated to overcome the difficulties of observing the galaxy from our position within the disk, using as tracers H II regions, open star clusters, FIR dust loops, or velocimetric deconvolution of CO and H I data (for a recent review see Vallée 2008). These made a fit to the data assuming 2, 4, or more symmetric, logarithmic spiral arms (e.g., Wainscoat 1992, Goncharov & Orlov 2003, Russeil 2003, Minchev & Quillan 2008, Gillman & Erenler 2008, Vallée 2008, etc.). Of course the resulting spiral-arm models feature symmetric spiral arms, because only those were allowed. This is neither proof nor evidence for the arms actually being symmetric. The velocimetric deconvolution of CO and H I has its own difficulties (Gomez 2006) and is conventionally performed assuming a purely circular flow of matter around the Galactic Center (e.g. Nakanishi & Sofue 2006). More complex flows are known to exist in the spiral arms themselves (e.g. Russeil et al. 2007), but also throughout the inner Galaxy, where the galactic bar induces a strong azimuthal asymmetry in the gravitational potential (Bissantz et al. 2003). Though nearly all current models of the Milky Way have four spiral arms, the method of fitting them varies widely. Most models force spiral arms into particular fits such as logarithmic arcs originating in the center of the galaxy (Vallée 2008). The model we will be using does not force-fit any presupposed pattern but rather is based on density enhancement of CO gas (Englmaier et al. 2008). This fixes more accurately and completely the position of the spiral arms especially on the far side of the Milky Way. This paper presents results of a re-examination of the previously found correlation between spiral arm crossings and the climate of Earth in light of new information on the Milky Way. It will be shown that due to the asymmetric position of the spiral arms, the correlation dissolves in this newer model and no periodic trend due to the passage of arms on less than the total orbital period is possible.

## 2. GALACTIC STRUCTURE

We will now examine the spiral-arm crossing locations in the more recent and complete view of the current galactic model. The models chosen take into consideration new criteria for spiral arms that do not force-fit them to logarithmic arcs (Englmaier et al. 2008). These models are based on the density of CO gas, which was recently modeled using a gas-flow model derived from smoothed particle hydrodynamics (SPH) simulations in gravitational potentials based on the NIR luminosity distribution of the bulge and disk (Bissantz et al. 2003). Besides providing a more accurate picture of cloud orbits in the inner Galaxy, a fundamental advantage of this model is that it provides kinematic resolution toward the Galactic center (Pohl et al. 2008), in contrast to standard deconvolution techniques based on purely circular rotation. To estimate the magnitude of systematic errors in the deconvolution, three different gas-flow models for the inner Galaxy were investigated, one of which was intentionally distorted so it no longer corresponds to a SPH simulation that has been adapted to gas data. A mismatch of the gas-flow model lead to a significant increase in the number of artifacts, thus allowing a quality ranking of models (for different bar parameters, etc.). In the inner Galaxy, where the CO line signal is strong enough and molecular clouds are presumed to trace the spiral arms, Pohl et al. (2008) note that two spiral arms seem to emerge at the ends of the bar, which have a small pitch angle. While some structures in the surface density distribution may be associated with two more arms, the evidence for those arms is not strong in their deconvolution. Englmaier et al. (2008) have compiled a composite model of the gas distribution in the Milky Way, using the CO distribution model for the inner Galaxy (galactocentric radii less than 8 kpc) where the SPH gas-flow model is available and CO line intensities are high, and a H I deconvolution from the literature (Levine et al. 2006) for the outer Galaxy, where the H I line provides a stronger signal and the asymmetry introduced by the galactic bar is small.
The resulting model of our Galaxy is shown in Fig 1. Since the spiral arms are matched to structure seen in the deprojected gas-distribution map, there is no reason to expect symmetry in the derived spiral pattern. However, one nevertheless finds an almost perfect 180-degree rotational symmetry in the inner Galaxy. At the solar circle the CO and H I deconvolutions match very well, thus permitting a sensible connection between the spiral arms in the inner and outer Galaxy, since arms cannot cross, only branch (Englmaier et al. 2008) thus find a 2-armed spiral pattern in the inner Galaxy, which branches in two more arms at about the solar circle, resulting in an asymmetric location of spiral arms. Such assymetry cannot be captured in the published studies, because they fit symmetric spiral-arm models. This gives us a better location for the Scutum-Crux arm, which is important for this work.

Although the speed of our solar system in its orbit around the galactic center is known to sufficient precision, the density-wave propagation speed of the spiral pattern is not. This pattern speed would have a dramatic affect on the intersection times we are modeling. Previous work on this subject has allowed the pattern speed to vary, and some studies suggest the co-rotation radius is close to the solar circle, meaning the spiral-arm pattern speed and the orbital angular velocity of our solar system may be very similar (e.g., Marochnik et al. 1972, Dias & Lépine 2005, etc.), in which case the period of spiral-arm passages would be much longer than the timescales of terrestrial climate change, 140 Myr. Gies & Helsel (2005) found extremely good fit to terrestrial variables for a difference between pattern speed and the orbital angular velocity of the solar system of 11.9 km s$^{-1}$ kpc$^{-1}$. They also considered a speed difference of 6.3 km s$^{-1}$ kpc$^{-1}$, but that does not agree as closely with the terrestrial variables, regardless of galactic model. Other work on the subject puts the difference in speeds at the value of 12.3 km s$^{-1}$ kpc$^{-1}$ (Svensmark 2006), and 11.1 km s$^{-1}$ kpc$^{-1}$ (Shaviv 2003). It appears that any speed difference between 1 and 13.5 km s$^{-1}$ kpc$^{-1}$ could be considered plausible (Shaviv 2003). Variations in the star-formation history of the Milky Way do not provide useful constraints on the spiral-arm pattern speed, because the time resolution is not high enough and because of orbit diffusion on Gyr-timescales (Rocha-Pinto et al. 2000).

The solar trajectory found by Gies & Helsel (2005) uses cylindrical coordinates and bases the current position of the solar system off of the model (model 2) developed by Dehnen & Binney (1998). This model places our sun starting at a distance of $R_0$=8.0 kpc and gives the solar system a circular velocity at that location of 217.4 km s$^{-1}$. From these starting parameters the trajectory was then numerically integrated backwards in time 500 Myr with time steps of 0.01 Myr to develop the elliptical orbit of our sun about the galactic center. As this path is elliptical, the solar motion relative will not be circular, but rather will vary in radius with a period less than the orbital period. Motion in the Z direction (normal to the galactic plane) is ignored as this will only slightly modify the intersection times and not change the pattern of intersections. The motion relative to the pattern is found by subtracting off the motion due to the assumed pattern speed. This assumes a constant pattern speed throughout the region defined by the variance in radius of the solar trajectory. Although this is assumed, velocity fields in this region based on the gas flow model of Bissantz et al. (2003) finds the velocity distortions to be less than 10 km s$^{-1}$, much less than the ~200 km s$^{-1}$ orbit velocity at the solar radius. Including the systematic errors in the gravitational potential, we estimate the error in the calculated solar trajectory to be below 10% and thus negligible in comparison with the uncertainty of the spiral-arm pattern speed. We also posit that all spiral arms have a common pattern speed. Freely assigning independent pattern speeds to the spiral arms introduces four free parameters (being the independent arm speeds), and in the light of only seven climate events to be reproduced simply violates Occam's razor. Intersections were then found as angles where the solar orbit and spiral arms coincide. These angles of intersection then correspond to the times given in Fig 2. Plotted against these points of intersection are vertical lines indicating the time of the last seven "Ice Age Epochs" (IAEs; Shaviv 2003). It should be noted at this time that the intersections are now quite asymmetric, mirroring the asymmetry present in our galactic picture. Whereas in previous work only periodic consequences of these crossings were considered, this new picture of the galaxy shows a lack of periodicity due to the asymmetry.

3. KEY CONNECTIONS TO TERRESTRIAL CLIMATE

Many different correlations between galactic position and the Earth's climate have been found, we will now re-examine those correlations in the light of the new galactic picture. For a record of Earth's climate we will both be using $\delta^{18}O$, as well as the times of the assumed IAEs over the last 1 Gyr. Two crucial linchpins of this work are a 140-Myr cycle apparent in both the oxygen data and the IAEs (Shaviv 2003, Shaviv & Veizer 2003). This strong 140-Myr cycle describes a massive temperature change unexplained as of yet by terrestrial causes. A similar and unexplained periodicity is also seen in biodiversity studies (Melott 2008; Melott & Bambach 2009), but has yet been seen to be statistically robust. The other important connection is the implied relatively recent transit of the Scutum-Crux spiral arm, linked to a probable cold period in the mid to late Jurassic (Shaviv 2003; Svensmark 2006).

We examine the path of the Sun relative to the spiral structure evident in Figure 1. The results are very different from those seen before. Though the solar motion now includes the elliptical path found by Gies & Helsel (2005), the motion is still very close to circular. The differences thus lie in the location of the spiral-arms in the new model and the new intersection angles.

4. RESULTS

As the new galactic structure is not symmetric, having dropped the former force-fit symmetry assumptions, it is impossible to produce a 140-Myr periodic spiral arm passage time. Figure 2 shows the timing of ice ages, the times of spiral arm passages found before, and those found with the new structural information. Horizontal error bars represent the time that the Sun's orbital path skims along inside or very close to one spiral arm. The only periodic trend that can be found with the new data is the relative orbital period of our solar system (Gies & Helsel 2005) relative to the previously assumed pattern speed around the galactic plane, which is slightly larger than 500 Myr. Though one could create varying periodic trends by changing this pattern speed, the orbital period relative to the galactic pattern could never reach the 140 Myr time as this is less than the orbital period itself, meaning the pattern and the Sun would be required to move in opposite directions.

We have checked the effect of varying the pattern speed in light of the elliptical orbit, and this does not introduce any improved fit to the geological data. Secondly, the Scutum-Crux spiral arm passage, key to fitting some of the most recent and reliable geological data, simply does not happen. The Sun never passes closer than 2 kpc to this spiral arm.

5. DISCUSSION

Although previous work found a correlation between the 140-Myr climate cycle on Earth and the intersection with spiral arms (Shaviv 2003, Shaviv & Veizer 2003, Svensmark 2006), with new data on the structure of the galaxy, this correlation disappears. We have used a new model of the large-scale gas distribution in the Galaxy, using a velocity-deconvolution of CO and H I line data that based on self-consistently computed, non-circular gas flows in the inner Galaxy (Bissantz et al. 2003, Pohl et al. 2008, Englmaier et al. 2008). In contrast to many published studies, this model does not force azimuthal symmetry into the spiral-arm structure. The asymmetry of the arms near the solar circle erases any correlation to the 140-Myr cycle and any periodic trend less than the orbital period of our solar system relative to the spiral pattern as a whole. This would be greater than 500 Myr for the previously fit pattern speed. Even if we allow the pattern speed to vary, it will not be less than the orbital period of the Sun, which is still longer than the 140 Myr cycle in question. The asymmetry of the new galactic picture could create a correlation between the spiral arm crossings and any non-periodic event by varying the pattern speed. We conclude that based on this new data there is no evidence to suggest any correlation between the transit of our solar system through the spiral arms of our galaxy and the terrestrial climate.

# 6. ACKNOWLEDGMENTS

ALM acknowledges support from NASA grant NNX09AM85G.

Support for MP by NASA grant NAG5-13559 is gratefully acknowledged.

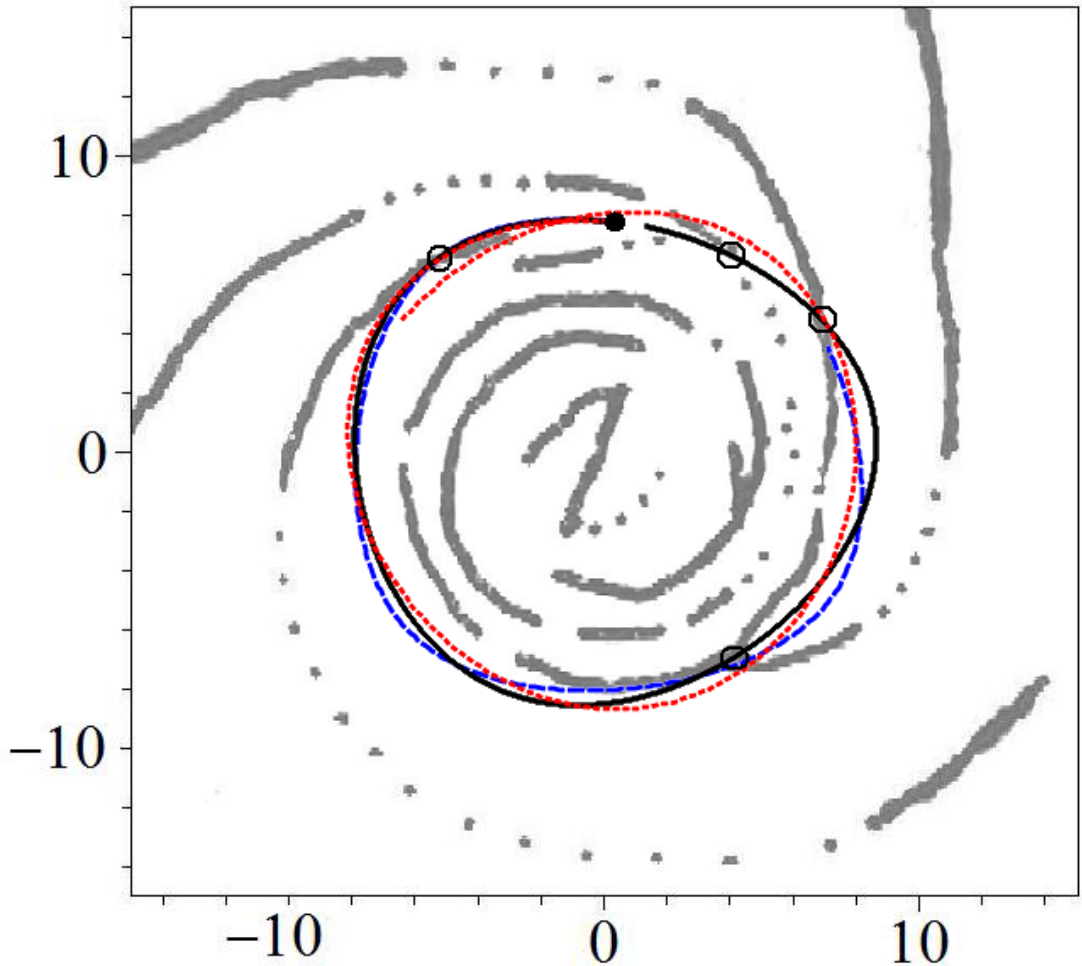

Figure 1: The path of our Solar System through the spiral arms of the Milky Way, as defined by high-density molecular-cloud data (Englmaier et al. 2008). The heavy black line is the trajectory of our solar system relative to the galactic pattern when the difference in speeds is 11.9 km s$^{-1}$ kpc$^{-1}$. The blue dashed line represents the lower bound of the previously assumed pattern speed difference, 10.1 km s$^{-1}$ kpc$^{-1}$; while the red dotted line represents the upper bound of the previously assumed pattern speed difference, 13.7 km s$^{-1}$ kpc$^{-1}$ (Shaviv 2003, Gies & Helsel 2005, Svensmark 2006). Our current location in the Galaxy is shown as a black dot, and the empty circles represent the intersections with the spiral arms. Axes are labeled in kpc.

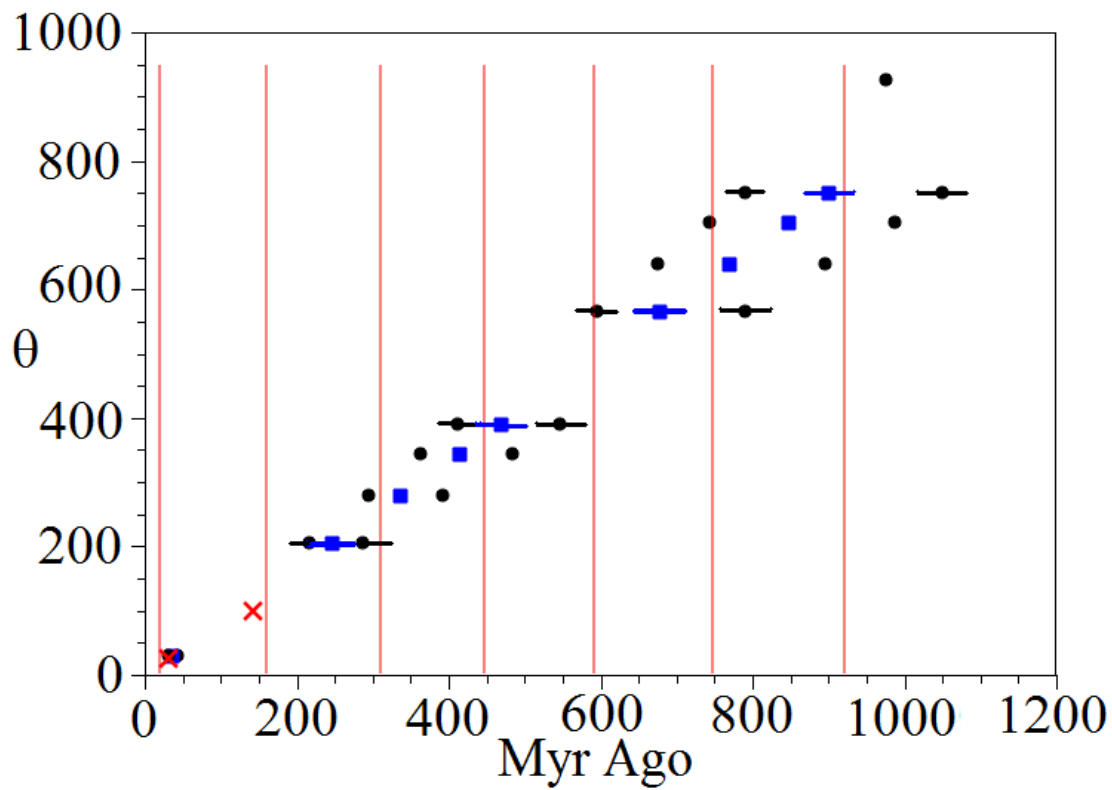

Figure 2: Red vertical lines represent the midpoints of the last seven ice ages (Shaviv 2003) showing the reported 140-Myr climatic cycle, red X's marking the intersections of the Solar System with spiral arms found in previous work (Svensmark 2006). Those were assumed to extend in a 140-Myr periodic pattern. Intersections of our Solar System with the spiral arms of the Milky Way computed from the current model are plotted in blue, with upper and lower bounds of pattern speed giving the intersections plotted in black, showing lack of correlation to the ice age epochs within the given pattern speed range. Horizontal error bars represent the extended intersection time for grazing events. Angles are taken in degrees from current location, increasing with time into the past.